\documentclass[aps,preprint,nofootinbib,groupedaddress,showkeys]{revtex4}

\usepackage{graphics}
\usepackage{graphicx}
\usepackage{amssymb}
\usepackage{amsmath}
\usepackage{amsfonts}

\newcommand{\vect}[1]{\boldsymbol{#1}}

\newif\ifFIG
\FIGtrue

\begin{document}

\title{Electronic stress tensor of the hydrogen molecular ion: comparison between the exact wave function and 
approximate wave functions using Gaussian basis sets}

\author{Kazuhide Ichikawa}
\author{Ayumu Wagatsuma}
\author{Masashi Kusumoto}
\author{Akitomo Tachibana} 
\email{akitomo@scl.kyoto-u.ac.jp}
\affiliation{Department of Micro Engineering, Kyoto University, Kyoto 606-8501, Japan}

\date{\today}

\begin{abstract}
We investigate the electronic stress tensor of the hydrogen molecular ion H$_2^+$ for the ground state 
 using the exact wave function and wave functions approximated by gaussian function basis set expansion. 
 The spatial distribution of the largest eigenvalue, corresponding eigenvectors, tension and kinetic energy density are compared. 
 We find that the cc-pV6Z basis set gives the spindle structure very close to the one calculated from the exact wave function. 
Similarly, energy density at the Lagrange point is very well approximated by the cc-pV5Z or cc-pV6Z basis sets.
 \end{abstract}




\maketitle

\section{Introduction}

There have been many studies of stress tensors in quantum systems as is found in Refs~\cite{Epstein1975, Bader1980, Bamzai1981a, Nielsen1983, Nielsen1985, Folland1986a, Folland1986b, Godfrey1988, Filippetti2000, Tachibana2001, Pendas2002, Rogers2002, Tachibana2004, Tachibana2005, Morante2006, Szarek2007, Tao2008, Ayers2009,Tachibana2010}, and even earlier \cite{Schrodinger1927,Pauli}. Several ways to define and apply the stress tensor in quantum systems are proposed in the literature. In this paper, we consider the electronic stress tensor and its application to chemical bonds and reactions as has been studied in Refs.~\cite{Tachibana2001,Tachibana2004,Tachibana2005,Szarek2007, Szarek2008, Szarek2009, Ichikawa2009b,Tachibana2010}. In these works, electronic stress tensors which measure effects caused by internal forces acting on electrons in molecules play a central role, in contrast to Ref.~\cite{Nielsen1983} and followers who focus on the stress tensor which is associated with forces on nuclei. 

Recently, the authors studied the electronic stress tensor of hydrogen molecular ion H$_2^+$ using its exact wave function \cite{Ichikawa2009a} (Refs.~\cite{Bamzai1981b,Godfrey1990} investigated stress tensor field of a H$_2^+$ molecule but with different definitions from ours). By using the exact wave functions, we were able to investigate the stress tensor and other associated quantities in a very accurate manner. 
However, since we can only obtain approximate wave functions for almost every molecular system, it is very important to study how the stress tensor of H$_2^+$ deviates from the exact form as we change the level of approximation of the quantum chemistry calculation to derive H$_2^+$ wave functions. This is the purpose of this paper. We believe this study helps to solidify the basis of the stress tensor analysis of molecular systems and open up areas of further application. 

This paper is organized as follows. In Sec.~\ref{sec:method}, we introduce quantities which are used in our stress tensor analysis. We also explain the exact and approximate wave functions we use. In Sec.~\ref{sec:result}, we compare the quantities obtained using the exact wave functions and approximate wave functions. Sec.~\ref{sec:conclusion} gives our conclusion. 

\section{Calculation Methods} \label{sec:method}

\subsection{Electronic stress tensor, tension and energy density} \label{sec:stress}

We first show our expression of the electronic stress tensor $\tau^{S}_{ij}$ constructed from the H$_2^+$ wave function $\psi$. That is,
\begin{eqnarray}
\tau^{S}_{ij} = \frac{\hbar^2}{4m} \left(   \psi^* \partial_i \partial_j \psi - \partial_i \psi^* \partial_j \psi  + c.c. \right), \label{eq:stress}
\end{eqnarray}
where $\{i,j\} = \{1, 2, 3\}$ denote spatial coordinates, $m$ is the electron mass and $c.c.$ stands for complex conjugate. 
The expression for many-electron systems, that is for almost every other molecules, is found in  Ref.~\cite{Tachibana2001}.
There, the stress tensor has been derived by field theoretic method which is applicable to many-particle systems. Here, for illustrative purpose, we derive it using the Schr\"{o}dinger equation for a single electron in H$_2^+$. We adopt the adiabatic approximation in this paper.

The time-dependent Schr\"{o}dinger equation reads
\begin{eqnarray}
i\hbar \frac{\partial \psi}{\partial t} = -\frac{\hbar^2}{2m} \nabla^2 \psi + V \psi, \label{eq:schrodinger}
\end{eqnarray}
where the potential energy $V$ is assumed to be real. It is well known that the continuity equation $\partial n / \partial t + \vect{\nabla} \cdot \vect{j} = 0$ 
holds by defining probability density $n = | \psi |^2$ and probability flux $\vect{j} = (\hbar/m) {\rm Im} (\psi^* \vect{\nabla}\psi )$. Now, the equilibrium equation for the electronic stress tensor 
is obtained from an equation of motion of $\vect{j}$. Namely, using Eq.~\eqref{eq:schrodinger}, time derivative of $\vect{j}$ can be written as 
\begin{eqnarray}
\frac{\partial \vect{j}}{\partial t} = -\frac{\hbar^2}{4m^2}\left( \vect{\nabla}\psi \cdot \nabla^2 \psi^*  - \psi^* \vect{\nabla}\nabla^2\psi + c.c.  \right)  
-\frac{|\psi|^2}{m} \vect{\nabla} V. \label{eq:djdt} \nonumber
\end{eqnarray}
For the steady state which is the case of molecular systems we consider, since $\partial \vect{j} /\partial t = 0$, we obtain the equilibrium equation
\begin{eqnarray}
  \partial_j \tau^{S}_{ij} + F_{L,i} = 0, \label{eq:equilibrium}
\end{eqnarray}
where $F_{L,i} = - |\psi|^2 \partial_i V$ is the Lorentz force generated by nuclei. For later convenience, we here define the first term in the equation as tension $F_{\tau,i} \equiv \partial_j \tau^{S}_{ij}$.
 From this equation, we see that $\tau^{S}_{ij}$ is the stress caused by purely quantum mechanical effects. 
The tension is written explicitly as
\begin{eqnarray}
\vect{F}_\tau = -\frac{\hbar^2}{4m}\left( \vect{\nabla}\psi \cdot \nabla^2 \psi^*  - \psi^* \vect{\nabla}\nabla^2\psi + c.c.  \right).  \label{eq:tension}
\end{eqnarray}
Since this force, which cancels the classical Lorentz force at each point in space (Eq.~\eqref{eq:equilibrium}), also expresses purely quantum mechanical effects, it is considered to carry some information of a chemical bond. Actually, in Ref.~\cite{Szarek2007}, the point where the tension vanishes is defined as ``Lagrange point"  and it is proposed that we may calculate energy density [defined below, Eq.~\eqref{eq:energy}] at that point to define bond order. This new definition turns out to carry very nice features of bond order \cite{Szarek2007,Szarek2008,Szarek2009}.

We now give our definition of energy density $\varepsilon_\tau$ constructed from the stress tensor.
In Ref.~\cite{Tachibana2001}, it has been proposed to define energy density $\varepsilon^S_\tau$ from the trace of the stress tensor as
\begin{eqnarray}
\varepsilon^S_\tau &\equiv& \frac{1}{2} \sum_i  \tau^S_{ii} \\
&=& \frac{\hbar^2}{8m} \left(   \psi^* \nabla^2 \psi - \vect{\nabla} \psi^* \cdot \vect{\nabla} \psi  + c.c. \right). \label{eq:energy}
\end{eqnarray}
Note that this definition gives correct total energy when integrated over the whole space and the virial theorem is applied.

We finally describe the kinetic energy density $n_T$, which is given by 
\begin{eqnarray}
n_T = -\frac{\hbar^2}{2m} \cdot \frac{1}{2} \left( \psi^* \nabla^2 \psi  + c.c. \right).   \label{eq:ene_kin}
\end{eqnarray}
Using kinetic energy density, we can divide the whole space into the region with $n_T > 0$, where the electron classical motion is allowed, $n_T < 0$, where the classical motion is forbidden, and $n_T = 0$, the boundary between them. These regions are respectively called the electronic drop region $R_D$, the electronic atmosphere region $R_A$ and the electronic interface $S$ \cite{Tachibana2001}. The $S$ can give the effective shape of the molecule and is therefore an important region in particular. 

The numerical calculation of these quantities are performed by the MRDFT code developed by our group \cite{MRDFTv3}.

\subsection{Exact and approximate wave functions}

We calculate quantities defined in Sec.~\ref{sec:stress} using the exact wave functions and approximate wave functions expanded in Gaussian basis set. We use the exact wave function which is computed in Ref.~\cite{Bates1953}. The calculations of the stress tensor and tension using this wave function have been reported by the authors in Ref.~\cite{Ichikawa2009a}. In the present paper, we focus on the ground state and fix the internuclear distance to be the equilibrium distance $R_e = 2.0$\,bohrs. (To be more precise, $R_e=1.9972$\,bohrs \cite{Schaad1970} but it does not make practical difference in our argument below.)

For the calculation of approximate wave functions, we use Gaussian 09 code \cite{Gaussian09} to perform the Hartree-Fock calculation with the following basis sets: STO-3G \cite{Hehre69}, 6-31G \cite{Ditchfield71}, 6-311G \cite{Raghavachari80b} and cc-pV$x$Z ($x=$ D, T, Q, 5 and 6) \cite{Dunning89,Peterson94}. For 6-31G, we also consider adding polarization and/or diffuse functions (6-31++G, 6-31G** and 6-31++G**) \cite{Hariharan73, Frisch84}, and similar for 6-311G (6-311++G, 6-311G** and 6-311++G**) \cite{Raghavachari80b,Frisch84}. The cc-pV$x$Z basis sets include polarization functions in their definition and we consider adding diffuse functions to each of them (aug-cc-pV$x$Z) \cite{Kendall92,Peterson94}. We use Cartesian functions (i.e. 6 d, 10 f, 15 g and 21 h functions) for all of the basis sets. 
Namely, we consider 19 types of basis set. We note that since H$_2^+$ is a one-electron system, the accuracy of the calculation depends only on the basis set.

\section{Results and discussion}  \label{sec:result}

In this section, we calculate the quantities introduced in the previous section for the exact and approximate wave functions of H$_2^+$. We compare among them and investigate how the results differ from those of the exact wave functions depending on the basis sets. In the following figures, we take the origin of the coordinate to be the midpoint of the two H nuclei and $z$-axis to be the internuclear axis. We locate H atoms at $(z, x)=(-1.0, 0.0)$ and $(1.0, 0.0)$. Although this system has the rotational symmetry around the $z$-axis and reflection symmetry for the $x=0$ plane, we sometimes draw figure for the regions with $x<0$ or $z<0$ to facilitate the intuitive understanding of chemical bond of H$_2^+$. 

\subsection{Electron density}

\begin{figure}[!h]
\begin{center}
\ifFIG
\includegraphics[width=17cm]{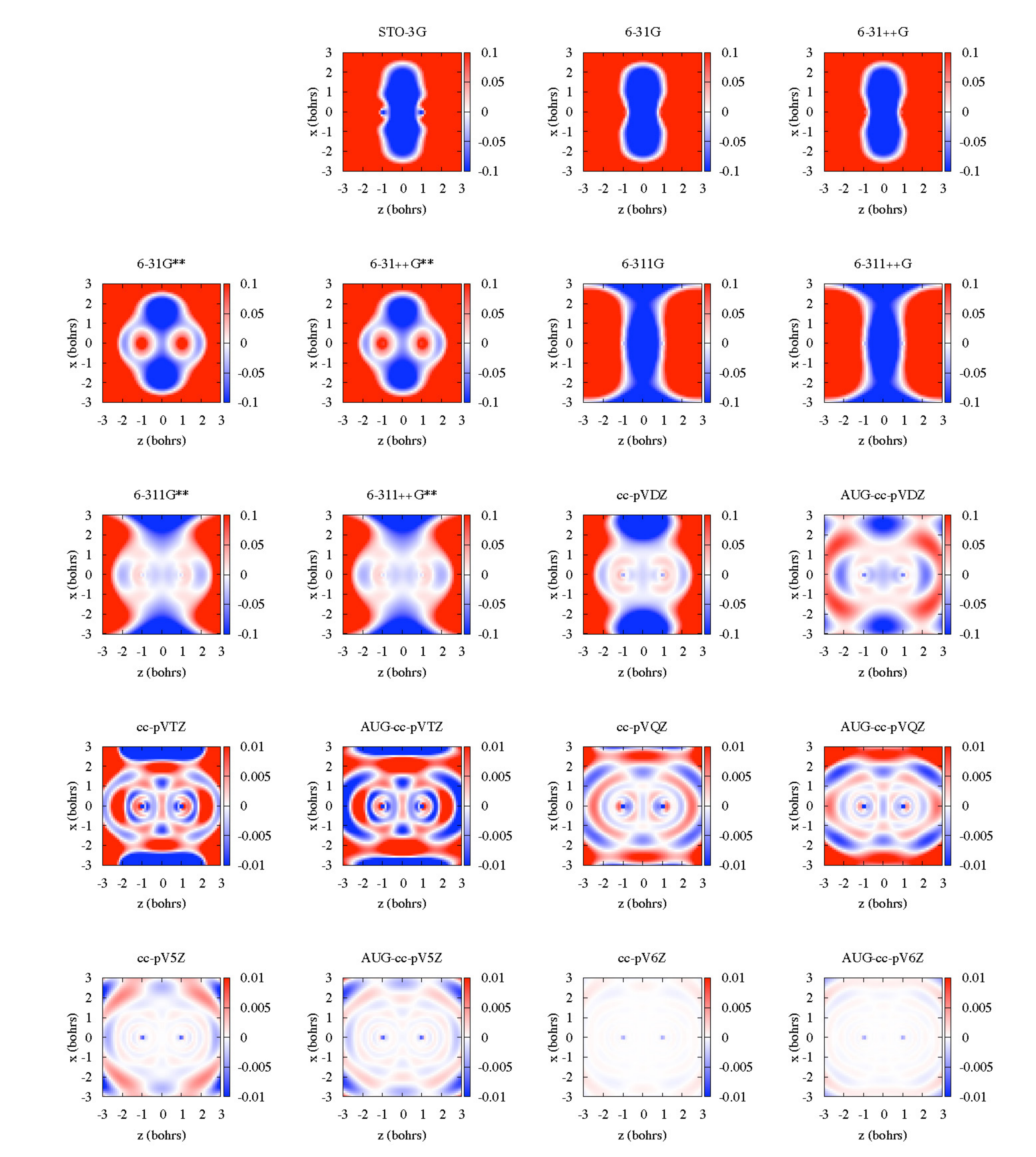}
\fi
 \caption{The spatial distribution of the relative error between the electron density calculated from the exact and approximate wave functions of H$_2^+$ molecule $\Delta n$ (Eq.~\eqref{eq:redens}) is plotted in the plane including two H nuclei (located at $(z, x) = (-1.0, 0.0)$ and (1.0, 0.0) ) for various basis sets. Note that the scale is ten times smaller after the cc-pVTZ basis set.}
 \label{fig:redens}
\end{center}
\end{figure}

Before we turn to discuss the electronic stress tensor, tension and so forth, it may be useful to see how the electron density varies with respect to the basis set we use. 
For this purpose, we plot the spatial distribution of relative error of the electron density $\Delta n$ in Fig.~\ref{fig:redens}. This is computed as
\begin{eqnarray}
\Delta n = \frac{n_{app} - n_{ex}}{n_{ex}},  \label{eq:redens}
\end{eqnarray}
where $n_{app}$ and $n_{ex}$ are the electron density from approximate wave functions and the exact wave function respectively. When we use cc-pV5Z or larger, the electron density is calculated with less than 1\% relative error in the region shown in the figure ($|x|<3$ and $|z|<3$). The best performance in this region is achieved by cc-pV6Z, which has 0.13\% error, rather than aug-cc-pV6Z, which has 0.44\% error. We can also confirm that adding polarization functions makes wave functions closer to the exact wave function especially in the region between nuclei whereas adding diffuse functions make little difference. 

\subsection{Eigenvalue and eigenvector of the stress tensor} 

\begin{figure}[!h]
\begin{center}
\ifFIG
\includegraphics[width=17cm]{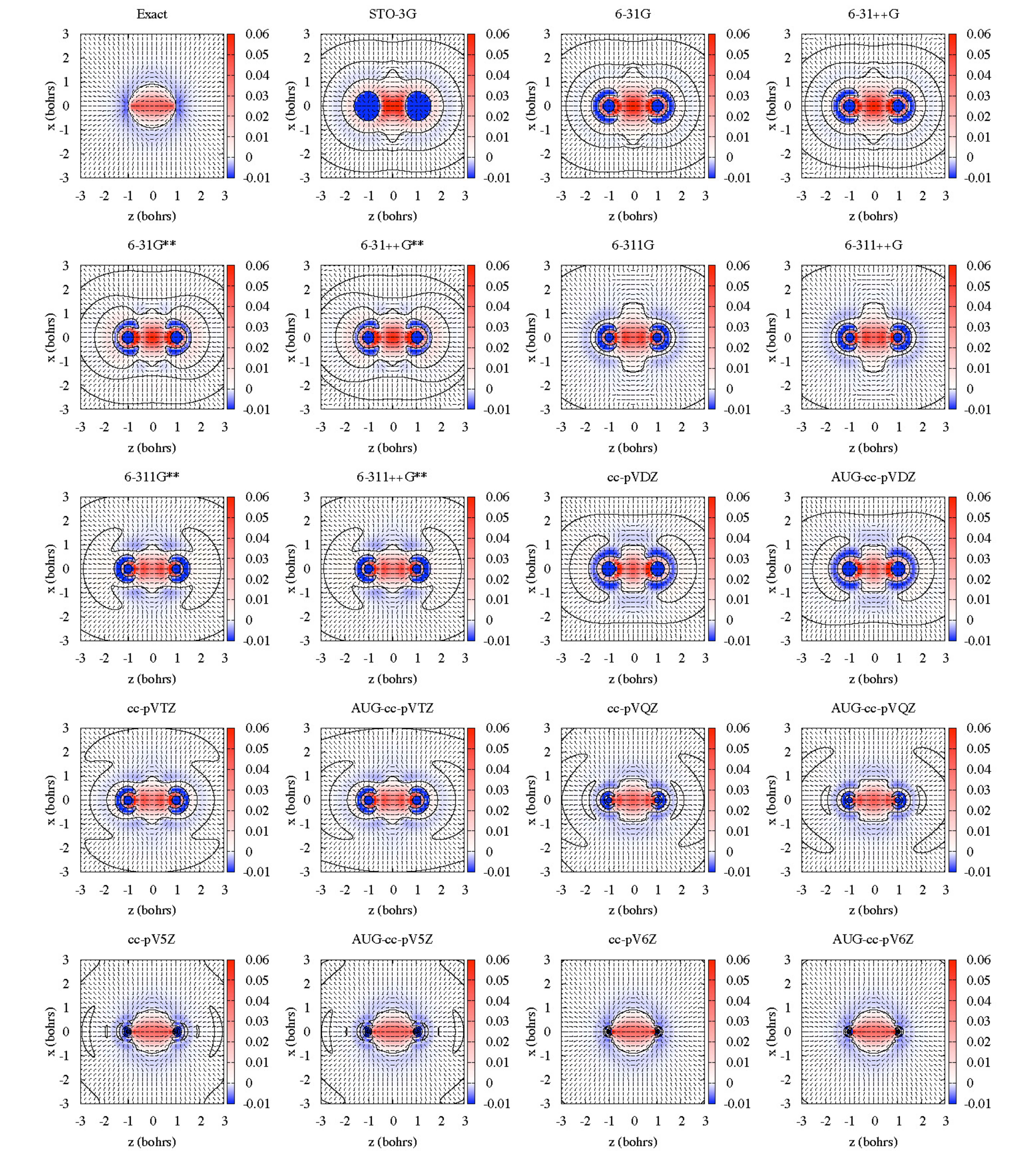}
\fi
 \caption{The spatial distribution of the largest eigenvalue and corresponding eigenvectors of the stress tensor of H$_2^+$ molecule is plotted in the plane including two H nuclei 
 (located at $(z, x) = (-1.0, 0.0)$ and (1.0, 0.0) ) for various basis sets. 
 The black solid line shows the zero surface of the eigenvalue.}
 \label{fig:eigvec}
\end{center}
\end{figure}

\begin{figure}[!h]
\begin{center}
\ifFIG
\includegraphics[width=17cm]{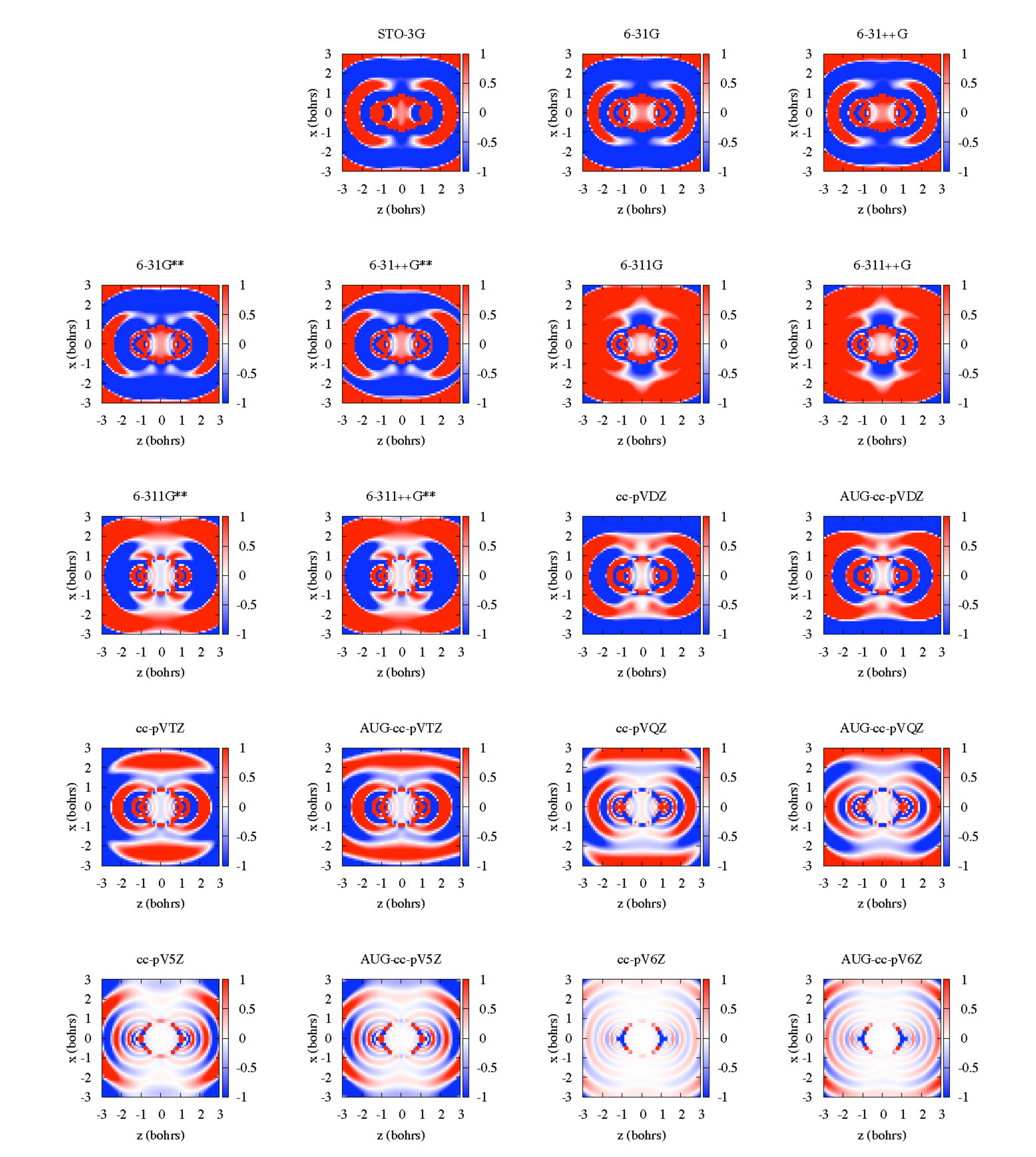}
\fi
 \caption{The spatial distribution of the relative error between the largest eigenvalue of the electronic stress tensor calculated from the exact and approximate wave functions of H$_2^+$ molecule $\Delta e$ (Eq.~\eqref{eq:reeig}) is plotted similarly to Fig.~\ref{fig:redens}. }
 \label{fig:reeig}
\end{center}
\end{figure}

\begin{figure}[!h]
\begin{center}
\ifFIG
\includegraphics[width=17cm]{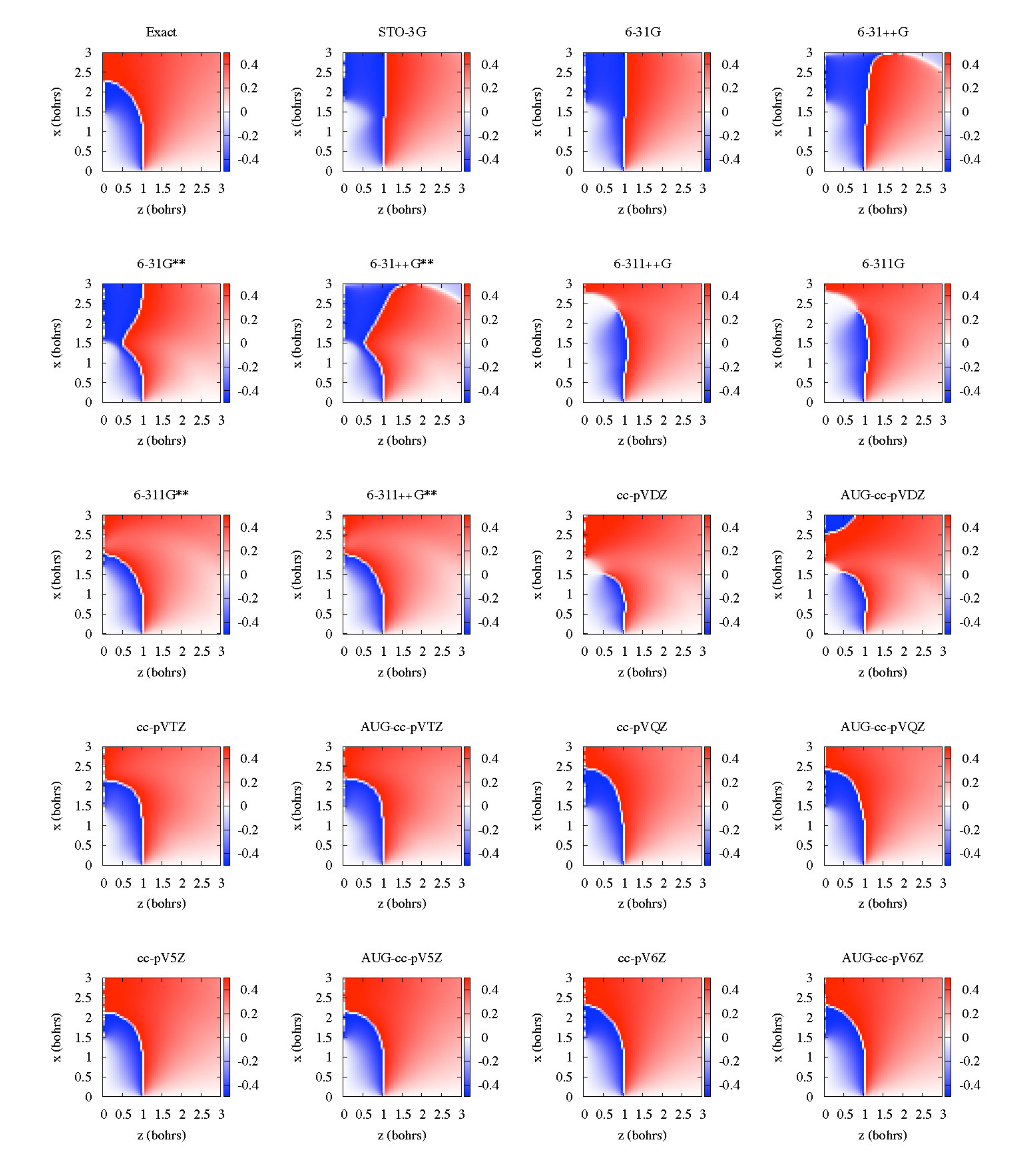}
\fi
 \caption{The spatial distribution of inclination angle $\theta_{eig}$ (Eq.~\eqref{eq:thetaeig}) of eigenvectors measured from the $z$-axis is plotted for ranges $x \ge 0$ and $z \ge 0$. $\theta_{eig}$ is shown in units of $\pi$\,radian and its range is $-\pi/2 < \theta_{eig} < \pi/2$.}
 \label{fig:atanvec}
\end{center}
\end{figure}

We here discuss the electronic stress tensor, defined by Eq.~\eqref{eq:stress}. As is often done for the stress tensor in general, we examine its largest eigenvalue and corresponding eigenvector. The sign of the largest eigenvalue tells whether electrons at a certain point in space feel tensile force (positive eigenvalue) or compressive force (negative eigenvalue) and the eigenvector tells direction of the force. They are shown in Fig.~\ref{fig:eigvec} for the exact wave function and approximate wave functions. In Fig.~\ref{fig:reeig}, we also plot the relative error of the largest eigenvalue $\Delta e$ 
\begin{eqnarray}
\Delta e = \frac{e_{app} - e_{ex}}{e_{ex}},  \label{eq:reeig}
\end{eqnarray}
where $e_{app}$ and $e_{ex}$ are the largest eigenvalues from approximate wave functions and the exact wave function respectively. 
To show the details of the direction of the eigenvector $\vect{v}_{eig}$, we plot 
\begin{eqnarray}
\theta_{eig} = \arctan \left( \frac{v_{eig,x}}{v_{eig,z}} \right),  \label{eq:thetaeig}
\end{eqnarray}
in Fig.~\ref{fig:atanvec}. This quantity ranges $-\pi/2 < \theta_{eig} < \pi/2$ and, in our $z$-$x$ plane, positive slope takes positive value (shown in red) and negative slope takes negative value (shown in blue). 
 
Let us first point out the properties found in the case of the exact wave function \cite{Ichikawa2009a}. The region of the positive eigenvalue occupies most of the space between the nuclei. 
In detail, the positive eigenvalue region is bounded by a closed sphere-like surface that touches two H nuclei. Also, in that region, the eigenvectors forms a bundle of flow lines that connects the H nuclei. This region, called ``spindle structure" \cite{Tachibana2004}, is clearly visible using the analysis with the exact wave function. 
Such structure expresses the correct directionality of the chemical bond and the positive eigenvalue of the stress tensor, implying a tensile stress, is considered to well characterize the covalent bond. 

As for the approximate wave functions, we see the region with positive eigenvalue (shown in red) spreads between the nuclei and also flows of the eigenvectors connecting the H nuclei. We can say that qualitative features, especially the spindle structure, are reproduced by the approximate wave functions. In detail, however, there are some differences. For one thing, the positive eigenvalue region in approximate cases surrounds the nuclei and it does not touch them (it is not simply connected in contrast to the case of the exact wave function). To put it another way, the positive region spreads outside the internuclear region and the negative region exists within the internuclear region. For the case of the exact wave function, there is only one zero surface of the eigenvalue and the whole space is divided into just a positive region and a negative region, but we see multiple zero surfaces for the case of the approximate wave functions in particular nearby and away from the nuclei. 

Comparing among the approximate wave functions, we see that the cc-pV6Z seems to quite closely reproduce the exact case. The cc-pV5Z also reproduces well the exact case. We may say that we need to approximate the wave function with less than 1\% error to have a good stress tensor profile. Similarly to the case of the electron density, the addition of the polarization functions improves the shape of the spindle structure whereas the diffuse functions do not. This is reasonable because the spindle structure is closely connected to the nature of bonding. 

It may be interesting to point out that the pattern of eigenvectors nearby the nuclei (that they point radially from the nuclei) 
 does not show much difference between the exact and approximate wave functions even for the basis sets which are not so large. Furthermore, such resemblance holds even at places where the signs of the eigenvalues are different between the exact and approximate cases. This is somewhat surprising feature and may be just a coincidence but we do not have an explanation. 

\subsection{Tension}

\begin{figure}[!h]
\begin{center}
\ifFIG
\includegraphics[width=17cm]{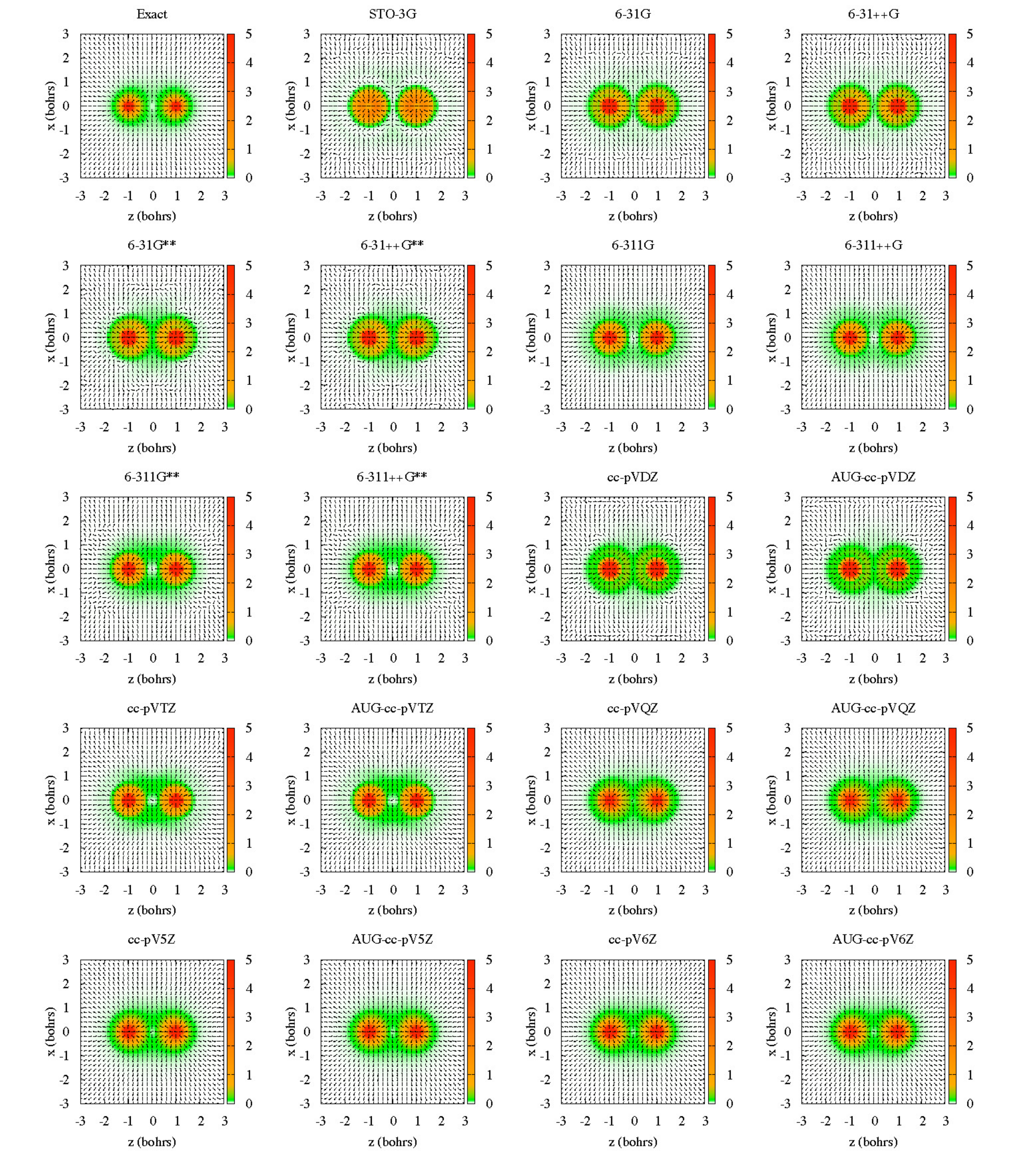}
\fi
 \caption{The spatial distributions of the tension $\vect{F}_\tau$ of H$_2^+$ molecule is plotted similarly to Fig.~\ref{fig:eigvec}. 
 Normalized tension vectors $\vect{F}_\tau/|\vect{F}_\tau|$ are shown by arrows and the norm $|\vect{F}_\tau|$ is depicted by a color map.}
 \label{fig:tension}
\end{center}
\end{figure}

\begin{figure}[!h]
\begin{center}
\ifFIG
\includegraphics[width=17cm]{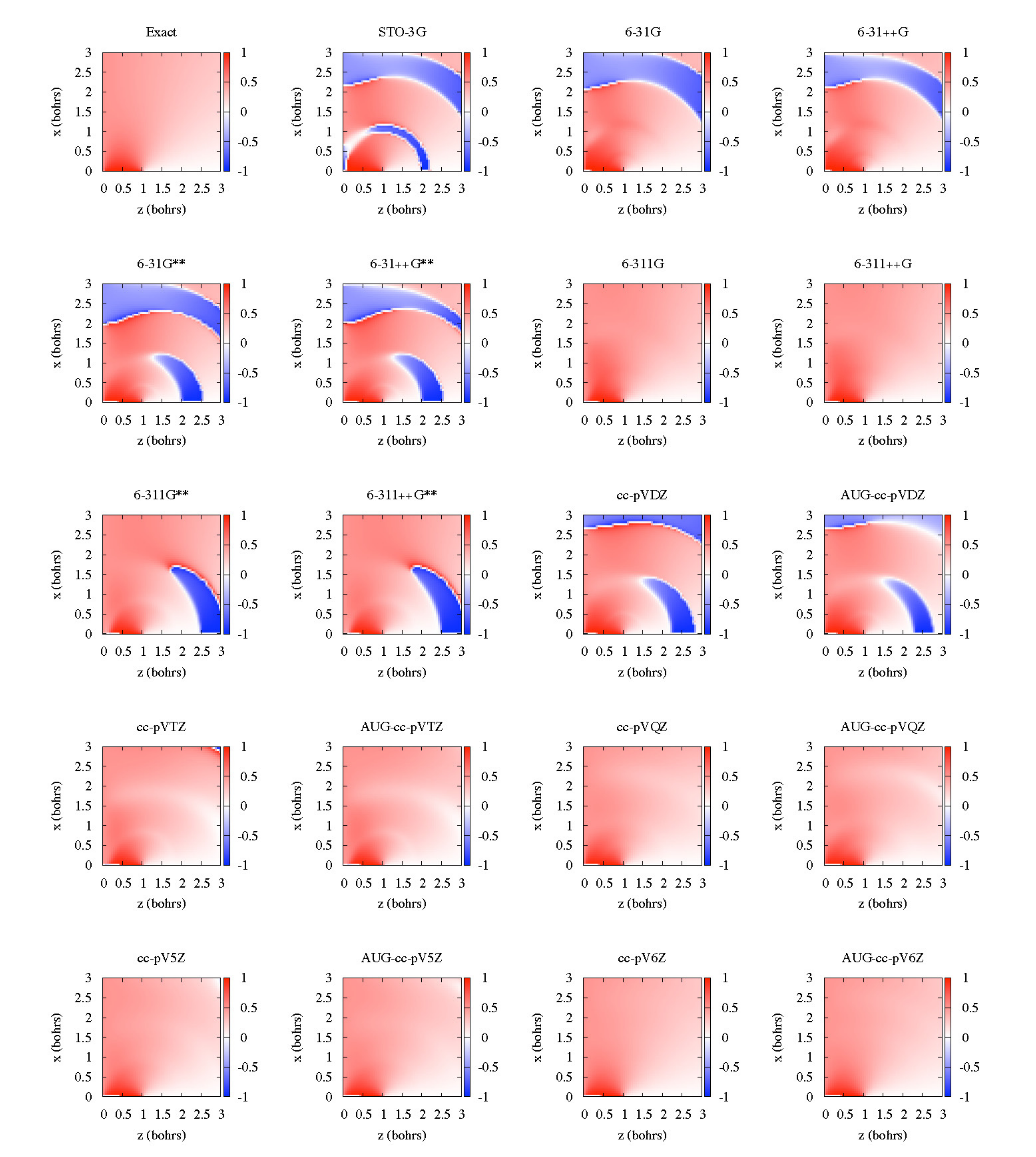}
\fi
 \caption{The spatial distribution of direction angle $\theta_{ten}$ (Eq.~\eqref{eq:thetaten}) of tension vectors measured from the $z$-axis is plotted for ranges $x \ge 0$ and $z \ge 0$. $\theta_{ten}$ is shown in units of $\pi$\,radian and its range is $-\pi < \theta_{ten} < \pi$.}
 \label{fig:atan2tension}
\end{center}
\end{figure}

Next, we examine the tension field computed as Eq.~\eqref{eq:tension}. This is in Fig.~\ref{fig:tension} where directions are shown by arrows and norm is shown by a color map. For the details of the direction, we plot 
\begin{eqnarray}
\theta_{ten} = {\rm atan2} \left(F_{\tau,x}, F_{\tau,z} \right),  \label{eq:thetaten}
\end{eqnarray}
in Fig.~\ref{fig:atan2tension}. 
\footnote{
This function can be written using the standard $\arctan$ function as follows:
\begin{eqnarray}
 {\rm atan2}(x,z) = \left \{
\begin{array}{l l }
\arctan (x/z) & \quad z > 0 \\
\pi + \arctan (x/z) & \quad x \ge 0, z<0  \\
-\pi + \arctan (x/z) & \quad x < 0, z<0 \\
\pi/2 & \quad x > 0 , z=0 \\
-\pi/2 & \quad x < 0 , z=0 \\
\end{array}
\right.
\end{eqnarray}
}
This is the angle between the positive $z$-axis and the point $(F_{\tau,z}, F_{\tau,x})$, with the range $-\pi < \theta_{ten} < \pi$. In our $z$-$x$ plane, the angle is positive for counter-clockwise angles (upper half-plane, $x>0$), and negative for clockwise angles (lower half-plane, $x<0$). 

When we compare the tension of the exact and approximate wave functions, it seems that 6-311G and cc-pV$x$Z ($x=$ D, T, Q, 5 and 6) do a good job in reproducing the exact result. In other words, compared with the case of the eigenvalue of the stress tensor, the tension can be computed rather accurately with smaller basis sets. This reflects the fact that the tension is just the opposite of the Coulomb force $\vect{F}_L = - |\psi|^2 \vect{\nabla} V$ (see Sec.~\ref{sec:stress}) and does not have much information as the stress tensor. 

The most important feature of the tension is the Lagrange points, where the tension vanishes, as mentioned in Sec.~\ref{sec:stress}. In the case of H$_2^+$, the Lagrange point turns out to be the midpoint of two H nuclei, $(0.0, 0.0)$. Actually, we see the region with small value of $|\vect{F}_\tau|$ (expressed in white) in the neighborhood of the origin for the exact wave function. Visual inspection of the figure tells us that such region is found in 6-311G, cc-pVTZ, cc-pV5Z and cc-pV6Z but not for cc-pVDZ and cc-pVQZ. Hence, relation between the area of the region with small $|\vect{F}_\tau|$ and the level of the basis set seems to be somewhat irregular. In the case of H$_2^+$, we can find a zero point in every basis set if we look at the data closely, but it may happen in general that the Lagrange points are found for smaller basis sets whereas not for larger basis sets. 

For the direction of the tension vector, as is found in the case of the eigenvector of the stress tensor discussed above, it is interesting that approximate wave functions give similar pattern to that of the exact wave function near the positions of the nuclei. 

\subsection{Kinetic energy density}

\begin{figure}[!h]
\begin{center}
\ifFIG
\includegraphics[width=17cm]{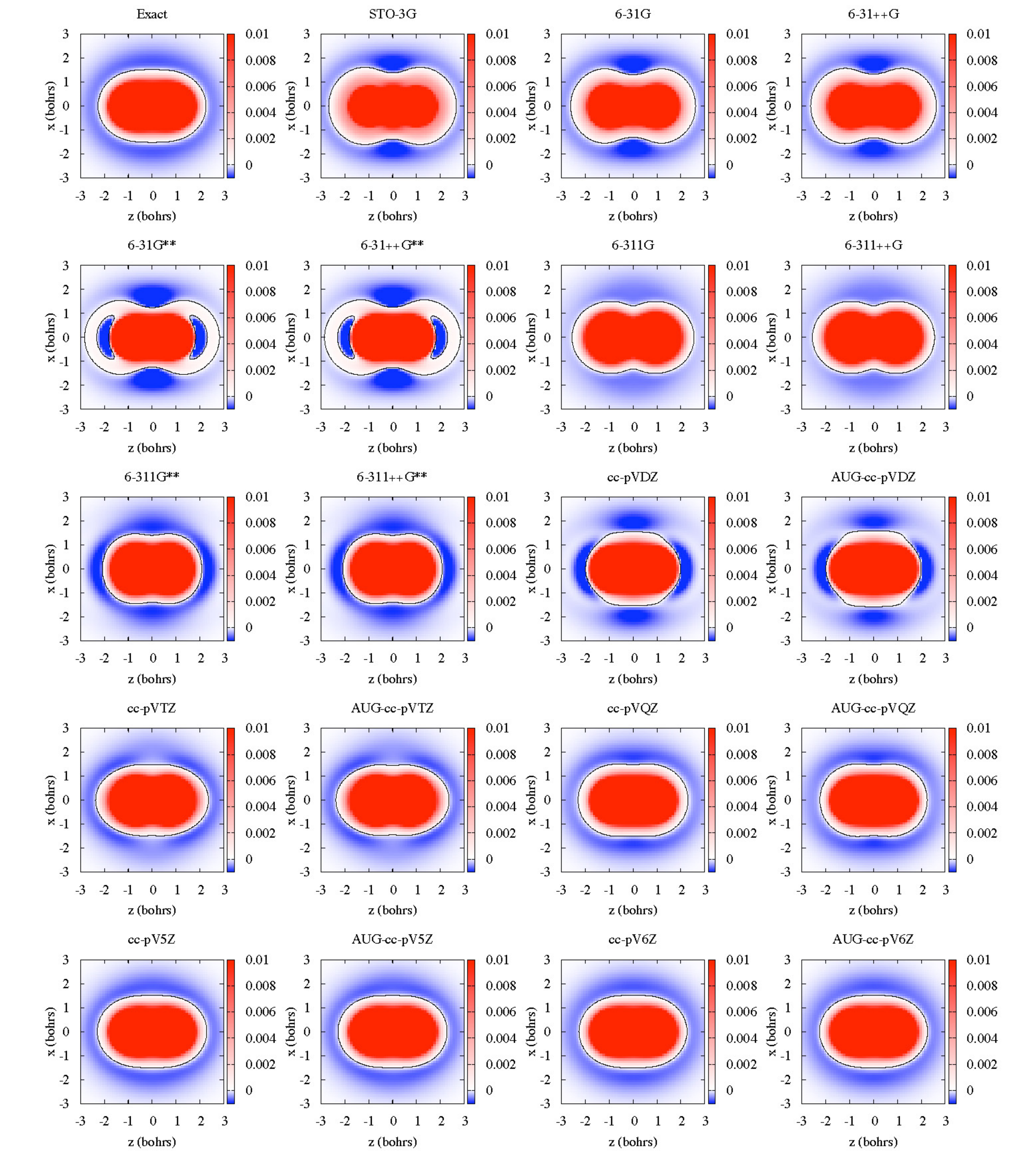}
\fi
 \caption{The spatial distribution of kinetic energy density is plotted similarly to Fig.~\ref{fig:eigvec}.  The black solid line shows the zero surface of the kinetic energy density.}
 \label{fig:enekin}
\end{center}
\end{figure}

We now examine the kinetic energy density defined as Eq.~\eqref{eq:ene_kin}. This is plotted in Fig.~\ref{fig:enekin}. The zero surface of the kinetic energy density can be considered as an effective surface of  a molecule and, for the exact wave function of the H$_2^+$ molecule, it is a closed surface which includes the H nuclei. This feature is reproduced by every approximate wave function but if we want to reproduce the detailed pattern of the curvature of the surface, it seems that we need cc-pV5Z or cc-pV6Z. 

In contrast to the case of the zero surface of the largest eigenvalue of the stress tensor discussed above, there is no extra zero surface around the nuclei for most of the basis sets (exceptions are 6-31G** and 6-31++G**).

\subsection{Quantities at Lagrange point}

As is mentioned in Sec.~\ref{sec:stress}, the Lagrange points are proposed to be useful to characterize chemical bond \cite{Szarek2007}. 
In particular, new bond order has been defined as the energy density (Eq.~\eqref{eq:energy}) at the Lagrange points \cite{Szarek2007,Szarek2008,Szarek2009}. 
Therefore, it is important to check the energy density at the Lagrange point is reproduced by approximate wave functions. 
We also compare the largest eigenvalue of the stress tensor.

\begin{figure}[!h]
\begin{center}
\includegraphics[width=17cm]{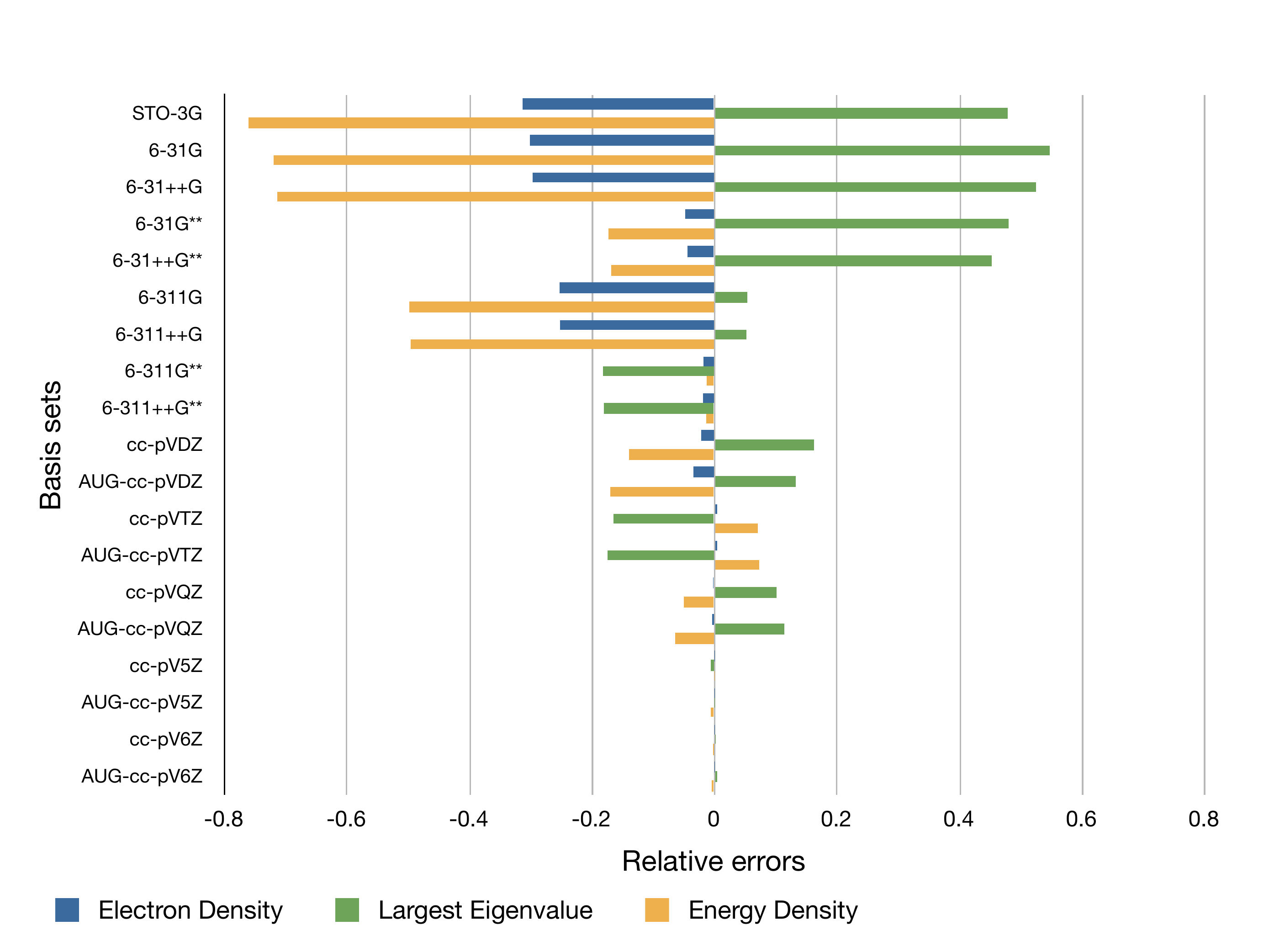}
 \caption{The relative error of the electron density, the largest eigenvalue of the stress tensor and energy density with respect to the exact wave function at the Lagrange point $(z, x)=(0.0, 0.0)$. See Table \ref{tab:reLPvalue} for the detailed numbers.}
 \label{fig:reLPvalue}
\end{center}
\end{figure}

In our case of the H$_2^+$ molecule, we have a Lagrange point at the midpoint of two H nuclei. We calculate the electron density, the largest eigenvalue of the stress tensor and energy density at the Lagrange point for the exact and approximate wave functions (Table~\ref{tab:LPvalue}) and also relative errors between those quantities derived from the exact and approximate wave functions (Table~\ref{tab:reLPvalue} and Fig.~\ref{fig:reLPvalue}). Roughly speaking, the largest eigenvalue and energy density at the Lagrange point are calculated with about 10 times larger relative error than that of electon density. We see that cc-pV5Z and cc-pV6Z can reproduce almost same results (less than 1\% relative error) as the exact wave function.

\section{Conclusion} \label{sec:conclusion}

In this paper, we investigated the electronic stress tensor of the hydrogen molecular ion H$_2^+$ for the ground state 
 using the exact wave function and wave functions approximated by gaussian function basis set expansion. 
We compared the spatial distribution of the largest eigenvalue of the stress tensor, corresponding eigenvectors, tension and kinetic energy density 
between the exact wave functions and approximate wave functions using 19 types of basis sets.
We also compared the energy density and the largest eigenvalue at the Lagrange point. 

 We found that the approximate wave function using the cc-pV6Z basis set gives the spindle structure, which is the key structure expressing the covalent bond,
 very close to the one calculated from the exact wave function. 
We also found that the energy density at the Lagrange point, which is the quantity used to define bond order, calculated by the exact wave function 
is very well approximated by the calculation using the cc-pV5Z or cc-pV6Z basis sets.

It is a good news that the electronic stress tensor and associated quantities can be computed accurately with commonly used basis sets which are 
stored in Gaussian program. Next step would be to compute the stress tensor for other molecules accurately and investigate their spindle structure in detail.



\newpage
\begin{table}
\begin{center}
\caption{Electron density, the largest eigenvalue of the stress tensor and energy density at the Lagrange point $(z, x)=(0.0, 0.0)$. All the values are multiplied by $10^2$.}
\label{tab:LPvalue}
\bigskip
\begin{tabular}{|r|c|c|c|}
\hline
Basis set & Electron density & Largest Eigenvalue & Energy density\\
 \hline 
 \hline
STO-3G &6.801&	5.943&	-1.063 \\
6-31G & 6.916&	6.219&	-1.246\\
6-31++G &6.960&	6.128&	-1.271\\
6-31G** &9.430&	5.947&	-3.674\\
6-31++G**& 9.463&	5.838&	-3.694\\
6-311G&7.398&	4.233&	-2.231\\
6-311++G&7.407&	4.228&	-2.238\\
6-311G** &9.725&	3.287&	-4.385\\
6-311++G** &9.720&	3.292&	-4.381\\
cc-pVDZ &9.685&	4.671&	-3.821\\
aug-cc-pVDZ &9.562&	4.552&	-3.688\\
cc-pVTZ &9.948&	3.356&	-4.756\\
aug-cc-pVTZ &9.943&	3.318&	-4.768\\
cc-pVQZ &9.880&	4.426&	-4.222\\
aug-cc-pVQZ &9.863&	4.477&	-4.156\\
cc-pV5Z &9.902&	3.995&	-4.445\\
aug-cc-pV5Z &9.898&	4.018&	-4.416\\
cc-pV6Z &9.902&	4.028&	-4.433\\
aug-cc-pV6Z &9.901&	4.037&	-4.423\\
\hline
 Exact & 9.903&	4.019&	-4.443 \\
   \hline
\end{tabular}
\end{center}
\end{table}

\begin{table}
\begin{center}
\caption{The relative error of the electron density, the largest eigenvalue of the stress tensor and energy density with respect to the exact wave function at the Lagrange point $(z, x)=(0.0, 0.0)$. The bar plot for this table is shown in Fig.~\ref{fig:reLPvalue}.}
\label{tab:reLPvalue}
\bigskip
\begin{tabular}{|r|c|c|c|}
\hline
Basis set & Electron density & Largest Eigenvalue & Energy density \\
 \hline 
 \hline
STO-3G &  -3.132$\times 10^{-1}$ &	4.787$\times 10^{-1}$ &	-7.607$\times 10^{-1}$\\
6-31G & -3.016$\times 10^{-1}$	& 5.473$\times 10^{-1}$	&-7.196$\times 10^{-1}$\\
6-31++G &-2.972$\times 10^{-1}$	&5.247$\times 10^{-1}$	&-7.139$\times 10^{-1}$\\
6-31G** &-4.776$\times 10^{-2}$	&4.797$\times 10^{-1}$	&-1.732$\times 10^{-1}$\\
6-31++G** &-4.441$\times 10^{-2}$	&4.526$\times 10^{-1}$	&-1.687$\times 10^{-1}$\\
6-311G & -2.530$\times 10^{-1}$	&5.325$\times 10^{-2}$	&-4.980$\times 10^{-1}$\\
6-311++G & -2.520$\times 10^{-1}$	&5.185$\times 10^{-2}$	&-4.964$\times 10^{-1}$\\
6-311G** & -1.800$\times 10^{-2}$	&-1.822$\times 10^{-1}$	&-1.300$\times 10^{-2}$\\
6-311++G** & -1.853$\times 10^{-2}$	&-1.810$\times 10^{-1}$	&-1.391$\times 10^{-2}$\\
cc-pVDZ &-2.200$\times 10^{-2}$	&1.622$\times 10^{-1}$	&-1.400$\times 10^{-1}$\\
aug-cc-pVDZ &-3.448$\times 10^{-2}$	&1.325$\times 10^{-1}$	&-1.699$\times 10^{-1}$\\
cc-pVTZ &4.537$\times 10^{-3}$	&-1.650$\times 10^{-1}$	&7.038$\times 10^{-2}$\\
aug-cc-pVTZ &4.004$\times 10^{-3}$	&-1.744$\times 10^{-1}$	&7.305$\times 10^{-2}$\\
cc-pVQZ &-2.323$\times 10^{-3}$	&1.012$\times 10^{-1}$	&-4.978$\times 10^{-2}$\\
aug-cc-pVQZ &-4.079$\times 10^{-3}$	&1.138$\times 10^{-1}$	&-6.461$\times 10^{-2}$\\
cc-pV5Z &-9.602$\times 10^{-5}$	&-5.961$\times 10^{-3}$	&3.359$\times 10^{-4}$\\
aug-cc-pV5Z &-5.210$\times 10^{-4}$	&-3.635$\times 10^{-4}$	&-6.084$\times 10^{-3}$\\
cc-pV6Z &-8.998$\times 10^{-5}$	&2.070$\times 10^{-3}$	&-2.292$\times 10^{-3}$\\
aug-cc-pV6Z &-1.910$\times 10^{-4}$	&4.384$\times 10^{-3}$	&-4.477$\times 10^{-3}$\\
  \hline
\end{tabular}
\end{center}
\end{table}

\newpage

\end{document}